\documentclass[osajnl,showpacs,superscriptaddress,10pt]{revtex4-1}
\usepackage{dcolumn}
\usepackage{bm}
\usepackage{multirow,array}
\usepackage{graphicx}
\usepackage{amsmath}
\usepackage{float}
\usepackage{xcolor}
\usepackage{empheq}

\usepackage{amsmath,amssymb,graphicx}
\begin{document}
\title{Quantum reflection of dark solitons scattered by reflectionless potential
barrier and position-dependent dispersion}

\author{L. Al Sakkaf, T. Uthayakumar, and U. Al Khawaja}\email{Corresponding author: u.alkhawaja@uaeu.ac.ae}
\affiliation{Department of Physics, United Arab Emirates University,
P.O. Box 15551, Al-Ain, United Arab Emirates}

\begin{abstract}
    We investigate theoretically and numerically quantum reflection of dark
solitons propagating through an external reflectionless potential
barrier or in the presence of a position-dependent dispersion. We
confirm that quantum reflection occurs in both cases with sharp
transition between complete reflection and complete transmission at
a critical initial soliton speed. The critical speed is calculated
numerically and analytically in terms of the soliton and potential
parameters. Analytical expressions for the critical speed were
derived using the exact trapped mode, a time-independent, and a
time-dependent variational calculations. It is then shown that
resonant scattering occurs at the critical speed, where the energy
of the incoming soliton is resonant with that of a trapped mode.
Reasonable agreement between analytical and numerical values for the
critical speed is obtained as long as a periodic multi-soliton
ejection regime is avoided.

\end{abstract}


\maketitle

\section{Introduction}
\label{introsec} One of the fascinating phenomena observed for
bright solitons in nonautonomous nonlinear systems is quantum
reflection. In such a phenomenon, the soliton that approaches a
potential can be reflected even in the absence of a classical
turning point \cite{well1,well2}. It portrays the wave nature of the
soliton. Furthermore, such nonclassical interactions are known to
exist only for the solitons approaching with lower velocities.
Whereas, if the incident soliton velocity is above a certain
critical value, a sharp transition from complete reflection to
complete transmission takes place. Subsequent investigations have
revealed that such a feature resuls from the formation of localised
trapped mode soliton at the centre of the potential. It also
signifies the indispensable role of the incident soliton energy
relative to the trapped mode energy on the appearance of sharp
transition of quantum reflection. Further, an accurate estimation of
the critical speed required for quantum reflection considering
various potential depths as well as investigating stability of the
trapped modes against perturbations for single as well as multi-node
trapped modes have also been studied \cite{usa}. Remarkably,
investigations from such systems described the possibility of the
high-speed soliton ejection, even for the stationary solitons
positioned at a distance sufficiently far from the centre of the
potential. These higher ejection speeds are accompanied with
multi-node trapped modes that hold larger binding energy
\cite{usa7}. Quantum reflection of solitons is witnessed in diverse
external potentials, for instance, barriers
\cite{bar0,bar1,bar2,bar3}, wells \cite{well1,well2,well3,well4},
steps \cite{step1,step2}, and surfaces \cite{surf1,surf2,surf3}.
Such studies allowed for the understanding of the energy exchange
mechanism of the soliton during the scattering process with the
potential to implement soliton based diodes, all-optical logic
gates, switches and filters using appropriate setups
\cite{usa1,usa2,usa3,usa4,usa5,usa6}.

In general, quantum reflection of bright solitons through diverse
potentials has been extensively analysed. For instance, in optics,
solitons are generated through compensating the group-velocity
dispersion (GVD) with the Kerr nonlinearity during its evolution
through an optical fiber. In view of nonlinear evolution with the
negative GVD (anomalous dispersion), the fiber supports the pulse
propagation in the form of sech pulse profile and such stable
nonlinear pulses are universally acknowledged as a bright soliton
\cite{Agrawal,Agrawal2,book}. Contrarily, for positive GVD (normal
dispersion), the soliton manifests itself as a localized dip in the
intensity with a uniform continuous-wave background. These dark
solitons also possess the shape and velocity preserving propagation
as that of its bright counterpart \cite{Hasegawa,Krokel,Kivshar}.
Since its recent theoretical prediction in optical fibers by
Hasegawa and Tappert \cite{Hasegawa} and experimental realization by
Emplit {\it et al.} \cite{Emplit}, considerable research efforts
have been made to understand the dynamics of dark solitons in
various situations under diverse nonlinear systems. In addition,
they are observed to be generated without the threshold in the input
pulse power \cite{Gredeskul}. Although dark solitons appear in
autonomous systems paramountly, their dynamics in nonautonomous
systems has particularly attracted a special interest to understand
the diverse real physical situations where complex space-time forces
appear \cite{Kivshar}.

Considering quantum reflection in dark solitons, Cheng, {\it et
al.}, demonstrated the quantum reflection in Bose-Einstein
condensates, where condensate comprises a dark soliton subjected to
the scattering by a potential barrier. This study discussed the
influence of diverse factors, namely, barrier height, width, the
orientation angle of the dark soliton and initial displacement of
the condensate cloud on the quantum reflection phenomenon. The
reflection probability is found to be sensitive to the initial
orientation and also influences the excitation process during the
condensate and barrier interaction \cite{Cheng}. A potential in the
form of box-like traps also accounted for the investigation of dark
soliton dynamics in BEC at zero temperature. Here, the sidewalls of
the potential are considered to rise either as Gaussian or a
power-law. In such a setup, the soliton is found to propagate
through the trap without any dissipation. However, dissipation was
observed during the reflection from a wall with the emission of
sound waves, resulting in a slight increase in the speed of the
solitons. For the multiple oscillations and reflections inside the
trap, the energy loss and the speed are found to increase
significantly \cite{Sciacca}. Scattering of dark solitons in a two
defect potential inferred that dark solitons cannot be trapped
through two identical potential barriers.  Moreover, interactions of
dark solitons in such a setup is found to increase the speed of the
soliton while traversing through the first barrier, which always
allows overcoming the second one. This predicted the only
possibility to achieve the optical diode based on dark soliton
scattering through time-dependent potential barriers
\cite{tsitoura}. An optical event horizon where reversible soliton
transformation from black soliton and a gray one to vice versa is
demonstrated through the interaction of dark soliton with a probe
wave. This reversible transformation of soliton is generated under
the competition between the internal phase of the dark
soliton and the nonlinear phase shift induced by the probe. Such a
system allows the potential application for information cloaking
\cite{Deng}.

The scattering of composite dark-bright solitons through a fixed
localized delta impurity is also considered by employing a mean-field approach. The study
identified that such interaction excited different modes that result
in the emergence of dark-bright soliton with a distinct velocity.
Also, accounted for the regions of reflection, transmission, and in
addition to the inelastic scattering behaviour with complex internal
mode excitations \cite{Majed}. Recently, Hansen {\it et. al},
described the propagation properties of matter wave solitons through
the localized scattering potentials and identified the regimes over
which solitons can behave as a wave or as a particle
as a consequence of the interplay between dispersion
and the attractive atomic interactions, through mean-field analysis
\cite{Hansen}. Although numerous investigations were reported for
dark solitons scattering in BEC, there is no significant study
involving quantum reflection of dark solitonic pulses through
scattering by potential barriers. In most of the situations, dark
solitons are dealt with and treated in presence of the bright soliton
background.

In the present study, we consider the scattering dynamics of dark
solitons propagated well far from the centre of an external
potential or a region of modulated dispersion in order to
investigate the existence and characteristics of quantum reflection.
The study has three main objectives: i) to confirm the existence of
quantum reflection and to identify the type of potentials (well or
barrier) needed for it to occur, ii) to investigate the
characteristics of the quantum reflection phenomenon in terms of the
main parameters of the system including the potential strength and
initial soliton size, and iii) to calculate both analytically and
numerically the critical soliton speed for quantum reflection.
Furthermore, our numerical investigations will show that for certain
parameters regime, a stimulated periodic multi-dark-soliton ejection
will be triggered upon the scattering. We begin with a derivation
considering both exact bright and dark soliton solutions that
deduces the necessary conditions required for quantum reflection of
dark solitons. Our study shows that the occurrence of quantum
reflection in dark solitons entails the potential in the form of a
barrier, which is in contrast with the situation for bright solitons
where it should be a potential well. We analyze the scattering
dynamics of dark solitons with two kinds of potential barrier
settings, namely, (i) with an external potential barrier, and (ii) a
position-dependent dispersion profile in the form of Dirac delta,
square well, sech, and sech$^2$ potential barriers. Here, the
position-dependent dispersion serves as an effective potential for
scattering.

We organize our present study as follows. In section \ref{funcsec},
we describe the scattering system and then derive the exact energy
functionals of both bright and dark solitons using the exact
solution of the nonlinear Schr\"odinger equation (NLSE), in section
\ref{exactsec}, and a variational calculation, in section
\ref{varsec}.  In section III, we calculate the effective potentials
in the presence of external potential and $x$-dependent dispersion.
Discussion of the necessary conditions required for the dark
solitons to be quantum reflected and numerical investigations of the
scattering dynamics is performed as well in this section. In section
IV, we use a time-dependent variational calculation to derive the
equations of motion. Finally, we summarize and discuss our main
findings in section V.

\section{Energy functional of the fundamental solitons}
\label{funcsec} In this section, we calculate the energy functional
for bright and dark solitons using the exact solutions of the NLSE.
We use also a variational calculation with an appropriate trial
function that leads to the exact energy functionals. The aim here is
to establish the notation and the theoretical framework. One of the
issues to settle at the outset is the divergency in the energy and
norm of the dark soliton due to its finite background. This
divergency is removed by shifting the intensity profile by its
asymptotic value at infinity.

We recall the fundamental bright and dark soliton solutions
supported by the well known fundamental NLSE, given by
\cite{Agrawal,Agrawal2,Agrawal2,book}
\begin{equation}
i\frac{\partial}{\partial t}\psi(x,t)=-g_1\frac{\partial^2}{\partial x^2}\psi(x,t)-g_2|\psi(x,t)|^2\psi(x,t)
\label{nlse},
\end{equation}
where  $\psi(x,t)$ is a complex function, $g_1$ and $g_2$ are
arbitrary real constants representing the strength of dispersion and
nonlinear terms, respectively. In nonlinear optics, the NLSE
describes the propagation of pulses in nonlinear media. In such a
context, the dispersion term corresponds to the group velocity
dispersion, which, depending on the sign of $g_1$, compresses or
spreads out the pulse, while the nonlinear term corresponds to the
Kerr effect, which describes the modulation of the refractive index
of the medium as a response to the propagating light pulse
intensity.

\subsection{Exact energy calculations of the fundamental  bright and dark solitons}
\label{exactsec}
In attractive nonlinear media, $g_2>0$, and normal dispersion, $g_1>0$, or alternatively repulsive nonlinear media, $g_2<0$, and anomalous dispersion, $g_1<0$, such that for both cases $g_2/g_1>0$, the NLSE, Eq. (\ref{nlse}), supports a movable bright soliton solution denoted by $\psi_B(x,t)$ and written as
\begin{equation}
\psi_B(x,t)=n\sqrt{\frac{{g_2}}{8 {g_1}}}\,
   \text{sech}\left[\frac{{g_2} n (x-{x_0}-v
   t)}{4 {g_1}}\right] e^{\frac{i}{16 g_1}\left[
   \left({g_2}^2 n^2-4 v^2\right) t+8 v
   \left(x-{x_0}\right)\right]}
   \label{bs},
\end{equation}
with a finite intensity normalised to $n$
\begin{eqnarray}\nonumber
N_B=\int_{-\infty}^{\infty}|\psi_B(x,t)|^2dx=n
\label{bsnorm},
\end{eqnarray}
 where $x_0$ and $v$ are the initial position and speed of the soliton centre. Energy of the  the bright soliton is given  by the energy functional
 \begin{eqnarray}\nonumber
 E_B&=&\int_{-\infty}^{\infty}\left[g_1\left|\frac{\partial}{\partial x}\psi_B(x,t)\right|^2-\frac{1}{2}g_2|\psi_B(x,t)|^4\right]dx\\
 &=&-\frac{g_2^2n^3}{48g_1}+ng_1v^2\label{bsenergy}.
 \end{eqnarray}
 In the contrary case where $g_2/g_1<0$, a movable dark soliton solution denoted by $\psi_D(x,t)$ exists and takes the following form
\begin{equation}
\psi_D(x,t)=n\sqrt{-\frac{{g_2}}{8 {g_1}}}\,
\text{tanh}\left[-\frac{{g_2} n (x-{x_0}-v
    t)}{4 {g_1}}\right] e^{-\frac{i}{8 g_1}\left[
    \left({g_2}^2 n^2+2 v^2\right) t-4 v
    \left(x-{x_0}\right)\right]}
\label{ds},
\end{equation}
 with a negative finite  intensity normalised to $-n$
\begin{eqnarray}\nonumber
N_D=\int_{-\infty}^{\infty}(|\psi_D(x,t)|^2-\rho_{\infty})dx=-n
\label{dsnorm},
\end{eqnarray}
where $\rho_{\infty}=\lim_{x\rightarrow\pm\infty} |\psi_D(x,t)|^2=-g_2n^2/(8g_1)$ is the background intensity.
The shift was necessary to avoid the divergency in the integral. The negative sign of the dark soliton norm is interpreted as a negative `mass' of a hole-like excitation. The negative value results from measuring the norm with respect to the finite background $\rho_{\infty}$.  To calculate the energy functional of dark soliton, a similar shift in intensity is needed in order to remove divergencies. This can be performed by expressing $\psi_D(x,t)$ in terms of intensity and phase as
\begin{equation}
\psi_D(x,t)=\sqrt{\rho(x,t)}\,e^{{i\phi(x,t)}}\label{dspolar}.
\end{equation}
Shifting the intensity as: $\rho(x,t)\rightarrow\rho(x,t)-\rho_{\infty}$, the energy of dark soliton then reads
\begin{eqnarray}
E_D&=&g_1\int_{-\infty}^{\infty}\left[(\rho (x,t)-\rho_\infty)\left(\frac{\partial}{\partial x} \phi(x,t)\right)^2+\left(\frac{\partial \sqrt{(\rho (x,t)-\rho_\infty)}}{\partial x}\right)^2\right]dx\nonumber\\
   &&-\frac{1}{2}g_2\int_{-\infty}^{\infty}(\rho (x,t)-\rho_\infty)^2dx.
\end{eqnarray}
Substituting for the $\rho(x,t)$ and $\phi(x,t)$ that correspond to solution (\ref{ds}), the energy functional takes the explicit form
\begin{equation}
E_D=\frac{g_2^2n^3}{48g_1}-ng_1v^2
\label{dsenergy},
\end{equation}\label{energyds}
which is equal to $-E_B$.

\subsection{Time-independent variational calculation for fundamental bright and dark solitons}
\label{varsec}
Here, we establish a variational calculation that reproduces the exact energy of bright and dark solitons. We perform a time-independent variational calculation using the trial functions
\begin{equation}
\psi_{B{\rm var}}=\sqrt{\frac{n\alpha}{2}}\,{\rm sech}[\alpha (x-x_0)]\,e^{iv (x-x_0)},
\label{bstrial}
\end{equation}
\begin{equation}
\psi_{D{\rm var}}=\sqrt{\frac{n\alpha}{2}}\,{\rm tanh}[\alpha (x-x_0)]\,e^{iv (x-x_0)}
\label{dstrial}.
\end{equation}
Using these trial functions, the energy expressions of both bright and dark soliton take the form
\begin{eqnarray}
E_{B{\rm var}}&=&n g_1v^2-\frac{1}{6}n^2g_2\alpha+\frac{1}{3}ng_1\alpha^2\label{nbvar2},\\
E_{D{\rm var}}&=&-n g_1v^2-\frac{1}{6}n^2g_2\alpha-\frac{1}{3}ng_1\alpha^2
\label{ndvar2}.
\end{eqnarray}
The above-mentioned intensity shift is applied here as well in order to obtain the dark soliton energy. The equilibrium value of the variational parameter is obtained by the condition $\partial E_{Bvar/Dvar}/\partial \alpha=0$, which gives $\alpha=ng_2/(4g_1)$ for the bright soliton and $\alpha=-ng_2/(4g_1)$ for the dark soliton. Substituting back into Eqs. (\ref{nbvar2}) and (\ref{ndvar2}) gives the energy expressions attained from the exact solutions (\ref{bsenergy}) and  (\ref{dsenergy}). An important difference between the bright soliton and dark soliton energy expressions should be noted. While the energy of the bright soliton has a minimum at the equilibrium value of $\alpha$, the dark soliton energy has a maximum. This indicates the stability of bright soliton and instability of dark soliton against shrinking or broadening of soliton width.

\section{NLSE with External potential and position-dependent dispersion}
\label{se2}
In the presence of an external potential, $V(x)$, or $x$-dependent dispersion, $f(x)$, the scattering dynamics of solitons is governed by the following NLSE
\begin{equation}
i\frac{\partial}{\partial t}\psi(x,t)=-g_1f(x)\frac{\partial^2}{\partial x^2}\psi(x,t)-g_2|\psi(x,t)|^2\psi(x,t)+V(x)\psi(x,t)
\label{nlse2},
\end{equation}
where  $\psi(x,t)$ is the field describing the intensity of the soliton. For matter-wave solitons in Bose-Einstein condensates,
it corresponds to the wave function of the condensate. For localised dispersion modulations, the $x$-dependent dispersion satisfies the boundary condition $\lim_{x\rightarrow\pm\infty} f(x)=1$,
which is guaranteed with the following form
\begin{equation}
f(x)=1+\delta(x)
\label{delta},
\end{equation}
where $\delta(x)$ is a localised function over a zero background.
Formally, a moving localised solution is written as
\begin{equation}\label{movsol}
\psi(x,t)=\sqrt{\rho(x-x_0-vt)}\,e^{iv(x-x_0)}.
\end{equation}
The energy functional corresponding to Eq. (\ref{nlse2}) reads
\begin{equation}
E=\int_{-\infty}^{\infty}\left[-g_1f(x)\psi^*(x,t)\frac{\partial^2}{\partial x^2}\psi(x,t)-\frac{1}{2}g_2|\psi(x,t)|^4+V(x)|\psi(x,t)|^2\right]dx,
\end{equation}
which takes the form
\begin{eqnarray}
E&=&\int_{-\infty}^{\infty}\left[g_1\left|\frac{\partial}{\partial x}\psi(x,t)\right|^2-\frac{1}{2}g_2|\psi(x,t)|^4\right]dx\nonumber\\
&&+g_1\int_{-\infty}^{\infty}\left[\delta(x)\left|\frac{\partial}{\partial x}\psi(x,t)\right|^2+\delta'(x)\psi^*(x,t)\frac{\partial}{\partial x}\psi(x,t)\right]dx\nonumber\\&&+\int_{-\infty}^{\infty}V(x)|\psi(x,t)|^2dx
\label{efunc},
\end{eqnarray}
where $(\cdot)'$ denotes a first derivative with respect to $x$ and $(\cdot)^*$ indicates complex conjugate. The first line in this equation is the energy of the fundamental soliton, the second and third lines correspond to the effective potentials resulting from the $x$-dependent dispersion and external potential, respectively. The energy is thus rewritten as
\begin{equation}\label{totenergy}
E(x_0,vt)=E_0+U_{disp}(x_0,vt)+U_{ext}(x_0,vt),
\end{equation}
 which upon using the moving localised solution, Eq. (\ref{movsol}), results in the following expressions
\begin{equation}\label{energy}
E_0=\int_{-\infty}^{\infty}\left[g_1v^2\rho(x-x_0-vt)+g_1\left(\frac{\partial \sqrt{\rho(x-x_0-vt)}}{\partial x}\right)^2-\frac{1}{2}g_2\rho^2(x-x_0-vt)\right]dx,
\end{equation}
\begin{equation}\label{disps0}
U_{disp}(x_0+vt,v)= g_1\left(v^2I_1+I_2+ivI_3\right),
\end{equation}
\begin{equation}\label{ext}
U_{ext}(x_0+vt)=\int_{-\infty}^{\infty} V(x) \rho (x-x_0-vt)dx,
\end{equation}
with
\begin{eqnarray}
I_1(x_0+vt)&=&\int_{-\infty}^{\infty}\delta (x) \rho (x-x_0-vt)dx,\\
I_2(x_0+vt)&=&\int_{-\infty}^{\infty}\frac{\partial \sqrt{\rho(x-x_0-vt)}}{\partial x}\left(\delta(x)\frac{\partial \sqrt{\rho(x-x_0-vt)}}{\partial x}+\delta'(x)\sqrt{\rho(x-x_0-vt)}\right)dx,\\I_3(x_0+vt)&=&\int_{-\infty}^{\infty}\delta
   '(x)\rho (x-x_0-vt) dx.
\end{eqnarray}
An important consequence of the $x$-dependent dispersion is the nonhermiticity of the hamiltonian indicated by the appearance of an imaginary part in the energy functional, $ivI_3(x_0+vt)$. However, the effect of this nonhermitian contribution is limited to the localised region of $\delta(x)$ where the scattering is taking place. At this region, the speed of the soliton is typically small which makes the effect of this term insignificant. Therefore, it can be neglected during the whole evolution time. From another perspective, we are interested in calculating the effective potential for a non moving initial soliton, $v=0$. Therefore, neither the first term nor the nonhermitian term will contribute to $U_{disp}$. The effective potential will be  then a function of only $x_0$, as follows
\begin{eqnarray}
U_{eff}(x_0)&\equiv&E(x_0)=E_0+U_{disp}(x_0)+U_{ext}(x_0).
\end{eqnarray}

Explicit formulae of $U_{eff}$ will be obtained below by considering specific forms of $V(x)$ and $\delta(x)$  for both bright and dark solitons, where again the replacement $\rho(x-x_0-vt)\rightarrow \rho(x-x_0-vt)-\rho_\infty$ is required for the latter to avoid divergency. For bright soliton, we employ the solution Eq. (\ref{bstrial}) which gives $\rho(x,t)=(n\alpha/2)\,{\rm sech}^2[\alpha (x-x_0)]$ and $\phi(x,t)=v(x-x_0)$. For the dark soliton, Eq. (\ref{dstrial}) gives $\rho(x,t)=(n\alpha/2)\,{\rm tanh}^2[\alpha (x-x_0)]$ and $\phi(x,t)=v(x-x_0)$ where the  shifted intensity is $\rho(x,t)-n\alpha/2=-(n\alpha/2)\,{\rm sech}^2[\alpha(x-x_0)]$.

\subsection{Effective external potential }
\begin{figure}[H]\centering
    \includegraphics[width=8cm,clip]{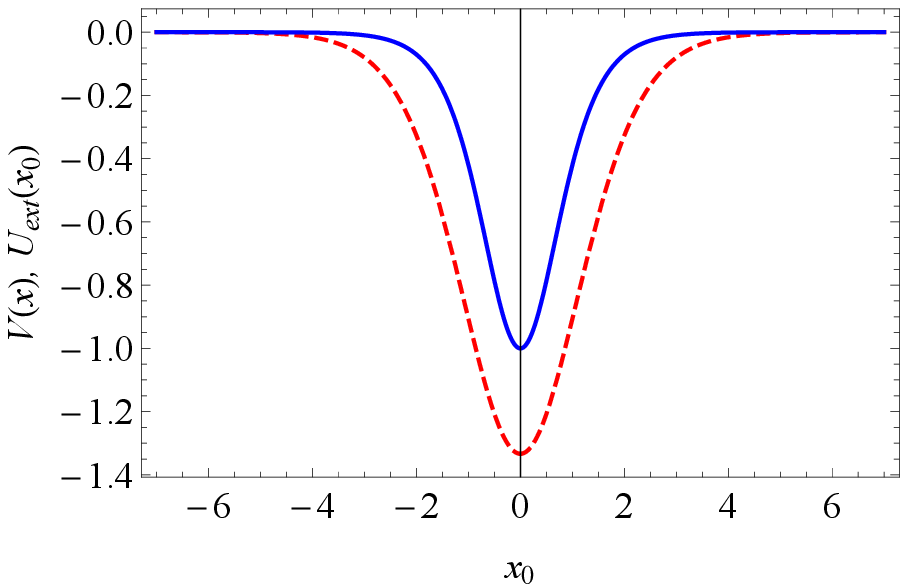}
    \begin{picture}(5,5)(5,5)
    \put(-24,130) {\color{black}{{\fcolorbox{white}{white}{\textbf{(a)}}}}}
    \end{picture}
    \includegraphics[width=8.11cm,clip]{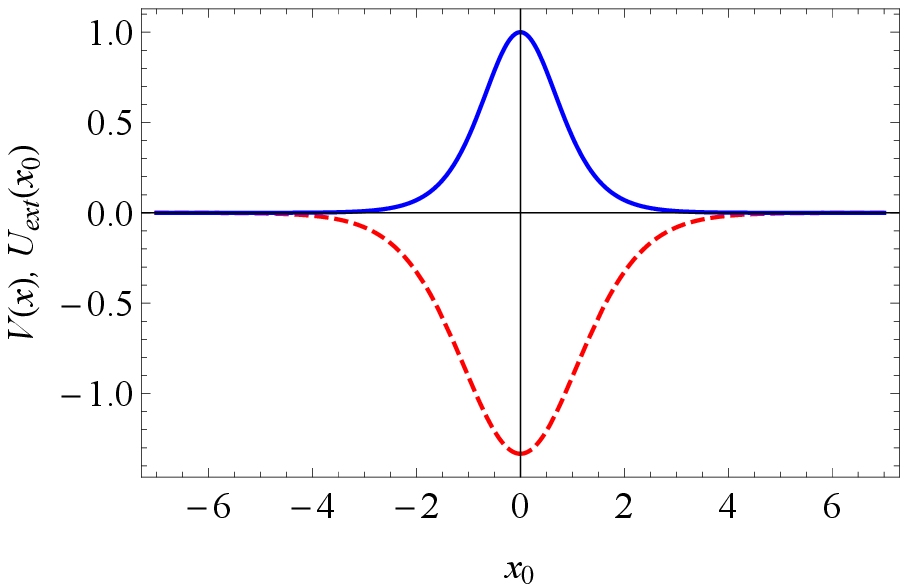}
    \begin{picture}(5,5)(5,5)
    \put(-27,130) {\color{black}{{\fcolorbox{white}{white}{\textbf{(b)}}}}}
    \end{picture}
    \caption{External potentials and effective external potentials as given by Eqs. (\ref{potential}) and  (\ref{effexter}),  (a) with bright soliton, $g_2=1$, and (b) with dark soliton, $g_2=-1$.  Solid blue and Dashed red curves correspond to the external potential and effective  potential, respectively. Parameters used are: $V_0=1,\,g_1=1/2,\,n=2$.}
    \label{fig1}
\end{figure}

For the scattering of dark solitons by an external potential, we consider the reflectionless P\"oschl-Teller potential provided by
\begin{equation}\label{potential}
V(x)=V_0\,{\rm sech}^2(\alpha x),
\end{equation}
where $V_0$ and $\alpha=\sqrt{|V_0|}$ are height and inverse width of potential. The corresponding effective external  potential can be obtained using Eq. (\ref{ext}), as
\begin{equation}\label{effexter}
U_{ext}(x_0)=\pm 2 n {V_0} \left[\alpha x_0 \coth (\alpha x_0)-1\right]
\text{csch}^2( \alpha x_0),
\end{equation}
where the positive and negative overall sign relate to the bright and dark solitons, respectively. This $\pm$ prefactor leads to an important conclusion about how the effective potential $U_{ext}(x_0)$ relates to the actual external potential $V(x)$. For a bright soliton, the effective potential will be a barrier (well) if the external potential is a barrier (well), but for the dark soliton the effective potential is a barrier (well) if $V(x)$ is a well (barrier). In Fig. \ref{fig1}, this is shown clearly where we plot the effective potential (\ref{effexter}) corresponding to the bright and dark solitons together with $V(x)$.

Quantum reflection takes place when the soliton is reflected by a region of effective potential well. Therefore, while quantum reflection of bright solitons takes place with $V(x)$ being a potential well, the situation is reversed for dark solitons. Quantum reflection of dark solitons takes place when $V(x)$ is a potential barrier.

In the following two subsections we investigate further the quantum reflection of a potential barrier both numerically and analytically. Specifically, we will show that, similar to bright solitons, sharp transition between full transmission and full reflection takes place at a critical soliton speed. The physics of quantum reflection turns out to be also similar to that of the bright soliton case where a trapped mode being formed at the potential. The critical speed for quantum reflection in terms of the strength of the potential $V_0$ will be also calculated using the numerical \cite{usa8} and analytical approaches.

\subsubsection{Calculation of the critical speed from the numerical solution of Eq. (\ref{nlse2})}
\begin{figure}[H]\centering
    \includegraphics[width=8cm,clip]{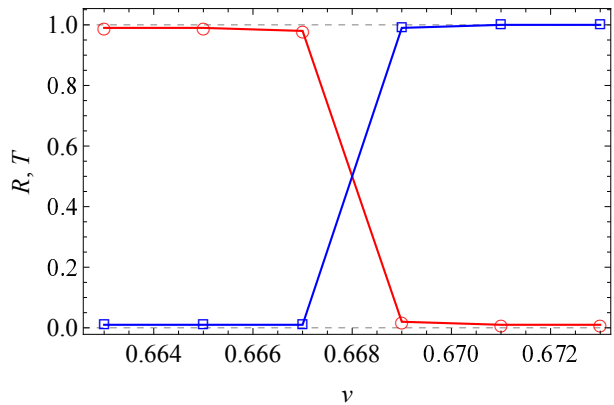}
    \caption{Reflection (red circles) and transmission (blue squares) coefficients versus initial speed of a dark soliton scattering by a reflectionless potential barrier Eq. (\ref{potential}). Parameters used are: $V_0=0.6$, $\alpha=\sqrt{V_0}$, $g_1=1/2$, $g_2=-1$, $x_0=-10$, $n=4$.}
    \label{fig2}
\end{figure}

\begin{figure}[H]\centering
\includegraphics[width=5cm,clip]{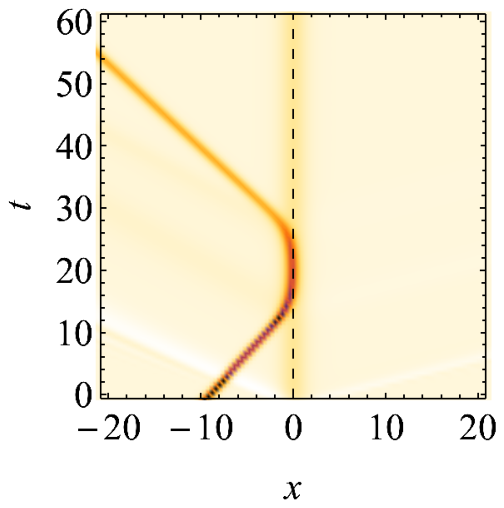}
\begin{picture}(5,5)(5,5)
\put(-28,132) {\color{black}{{\fcolorbox{white}{white}{\textbf{(a)}}}}}
\end{picture}
\includegraphics[width=5cm,clip]{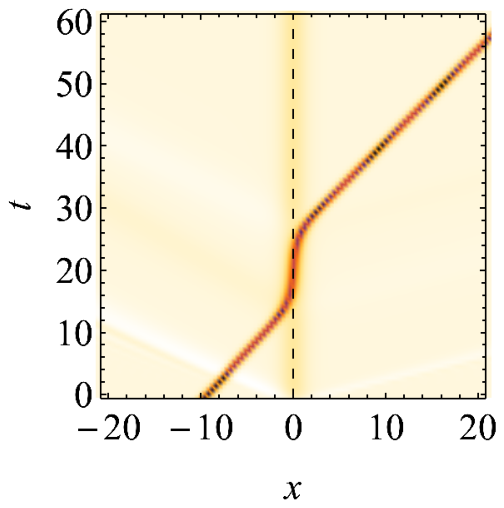}
\begin{picture}(5,5)(5,5)
\put(-105,132) {\color{black}{{\fcolorbox{white}{white}{\textbf{(b)}}}}}
\end{picture}
\includegraphics[width=5cm,clip]{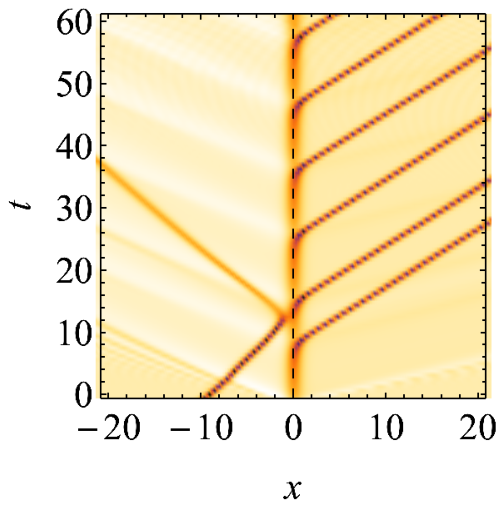}
\begin{picture}(5,5)(5,5)
\put(-105,132) {\color{black}{{\fcolorbox{white}{white}{\textbf{(c)}}}}}
\end{picture}
\caption{Quantum scattering of the dark soliton Eq. (\ref{ds})  by an effective potential Eq. (\ref{effexter}). The  external potential is indicated by the vertical dashed line. (a) Quantum reflection with critical speed $v=v_c=0.665$,  (b) transmission with $v=0.67$, and (c) multi-ejection $v=0.8$, $V_0=2$. Other parameters used are the same of those in Fig. \ref{fig2}.}
\label{fig3}
\end{figure}

\begin{figure}[htb]
\includegraphics[width=8cm]{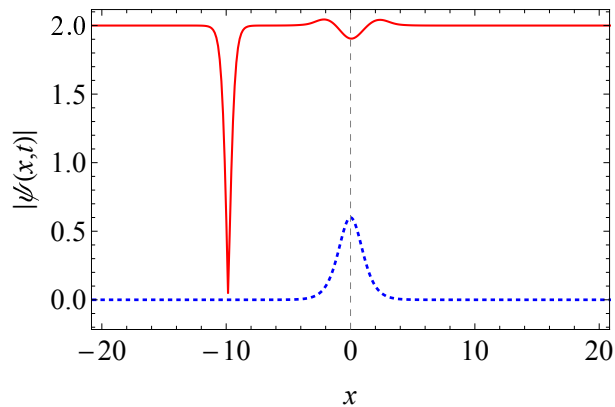}
 \caption{Existence of a trapped mode at the centre of the potential before the propagation. Dotted blue curve corresponds to the potential barrier which behaves as an effective potential well. Parameters used are: $V_0=0.6, \alpha=\sqrt{V_0},\,g_1=1/2,\,g_2=-1,\,n=4, v = 0.51$. }
  \label{fig4}
\end{figure}

We solve numerically Eq. (\ref{nlse2}) in the presence of the  external potential (\ref{potential}), where we set $\delta(x)=0$ and use $\psi_D(x,0)$ from Eq. (\ref{movsol}) as an initial profile. We define the scattering coefficients as
\begin{equation}
R=(1/N_D)\int_{-\infty}^{-l}(|\psi_D(x,t_f)|^2-\rho_{\infty})\,dx,
\label{r}
\end{equation}
\begin{equation}
T=(1/N_D)\int_{l}^{\infty}(|\psi_D(x,t_f)|^2-\rho_{\infty})\,dx,
\label{t}
\end{equation}
where $N_D$ is defined by Eq. (\ref{dsnorm}),  $R$ and $T$ are the scattering coefficients of reflection and transmission, respectively, $l$ is a length larger than the width of the potential, and $t_f$ is an evolution time such that the scattered soliton is sufficiently far from the potential.  Figure \ref{fig2} shows a clear possibility of obtaining quantum reflection where a sharp transmission of the coefficients from complete reflectance to complete transmittance is achieved at $v = 0.67$. This is confirmed by the spatio-temporal plots in Fig. \ref{fig3} (a) and (b), where the two different scattering outcomes are obtained for initial soliton speeds just below and above the critical speed.
The excitation of trapped mode is also verified in Fig. \ref{fig4}, where a snapshot shows a dark soliton at the potential centre. Higher energy trapped modes were also found for larger potential barrier strengths. However, in such a case, quantum reflections are found to be accompanied with a considerable amount of radiation, or even completely replaced by multi-soliton ejections, as indicated by Fig. \ref{fig3} (c).

\begin{figure}[H]\centering
    \includegraphics[width=8cm,clip]{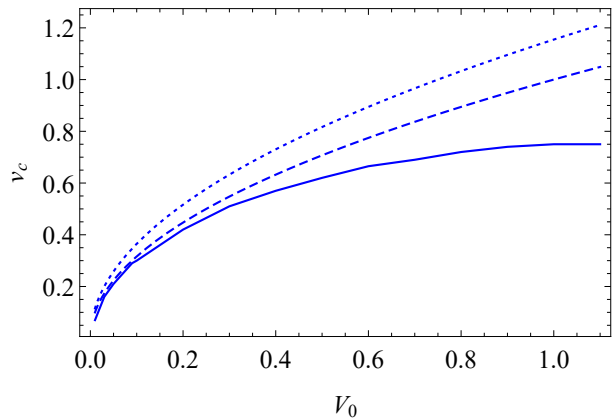}
    \caption{Critical soliton speed, $v_c$, at which the dark soliton will be
        reflected by the effective potential (\ref{effexter}) in terms of its strength, $V_0$. Solid curve is obtained from the numerical solution of Eq. (\ref{nlse2}), as explained in the text. Dashed curve is the analytical result (\ref{trapspeed2}). Dotted curve is the result of time-dependent variational calculation (\ref{cspeed1}).  Parameters used are: $\alpha=\sqrt{V_0}$, $g_1=1/2$, $g_2=-1$, $x_0=-10$, $n=4$.}
    \label{fig5}
\end{figure}

The dependence of the critical speed for quantum reflection on the strength of the potential is shown by the solid blue curve in Fig. \ref{fig5}. For $V_{0}>1.1$, the scattering of dark solitons results in the above-mentioned multi-soliton ejection behaviour as illustrated in Fig. \ref{fig3}(c). Hence, we restricted ourselves within the goal of investigating quantum reflection while the multi-soliton ejection will be considered for the study in near future.

\subsubsection{Calculation of the critical speed with exact trapped mode solution}
Based on our understanding to the mechanism of quantum reflection, a trapped mode plays a crucial role in determining the sharp transition between full transmission and full (quantum) reflection. Simply, if the energy of the incoming soliton is larger (less) than the energy of the trapped mode, the soliton will be fully transmitted (reflected). If the energy of the incident soliton is equal to that of the trapped mode, the soliton will be trapped, though such a trapped state is unstable against small perturbations in position or width of the soliton. Consequently, determining the properties of the trapped mode, mainly its profile and energy, are essential for accounting to the critical speed theoretically. Fortunately, an analytic exact solution, which is the nodeless trapped mode, exists for the NLSE with the reflectionless P\"oschl-Teller potential. Here, we will exploit this solution to present an analytic derivation of the critical speed in terms of the potential strength which will provide an independent account for our earlier numerical calculation which we can compare with.\\

In the absence of the $x$-dependent dispersion and with the presence of the P\"oschl-Teller potential, $V(x)=V_0\,{\rm sech}^2(\sqrt{V_0} x),$ the NLSE (\ref{nlse2}) with $g_1=1/2$, $g_2<0$, and $V_0>0$ admits the exact dark soliton solution
\begin{equation}\label{trapsoliton}
\psi_{trap}(x,t)=\sqrt{-\frac{2V_0}{g_2}}\,{\rm tanh}\left(\sqrt{V_0} x\right)e^{-2i V_0t},
\end{equation}
which upon substituting in the energy functional (\ref{totenergy}) with $U_{disp}=0$ and taking in to account the corresponding shifted intensity as $\rho(x,t)+2V_0/g_2=(2V_0/g_2)\,{\rm sech}^2(\sqrt{V_0} x)$, gives the exact trapped energy that takes the form
\begin{equation}\label{trapenergy}
E_{trap}=\frac{2V_0^{3/2}}{3 g_2}.
\end{equation}
The  initial profile of the soliton is taken as the exact dark soliton of the fundamental NLSE, namely (\ref{ds}), thus the  energy
of the initial soliton takes the form of  (\ref{dsenergy}). Equating the two energies in (\ref{dsenergy}) and (\ref{trapenergy}), $E_D=E_{trap}$, yields the exact  critical speed for quantum reflection
\begin{eqnarray}\label{trapspeed}
v_c=\sqrt{\frac{g_2^2n^2}{12}-\frac{4V_0^{3/2}}{3ng_2}}.
\end{eqnarray}
In addition to equating their energies, the norms of the initial and trapped solitons should be equal. As given by (\ref{dsnorm}), the norm of the initial soliton is $-n$. The norm of the trapped mode (\ref{trapsoliton}) is calculated to be $n=4\sqrt{V_0}\,/g_2$. Equating the two norms and then substituting for $n$ in Eq. (\ref{trapspeed}), we get
\begin{eqnarray}\label{trapspeed2}
v_c=\sqrt{V_0}.
\end{eqnarray}
This theoretical result agrees favourably with the numerical calculation, as shown by dashed blue curve in Fig. \ref{fig5}, especially for smaller values of $V_0$. As noted above, for larger values of $V_0$, radiation increases and keeping in mind that the theoretical calculation of $v_c$ does not take into account radiated energy, this explains the increased deviation of the theoretical result from the numerical one.
\begin{figure}[H]\centering
    \includegraphics[width=5cm,clip]{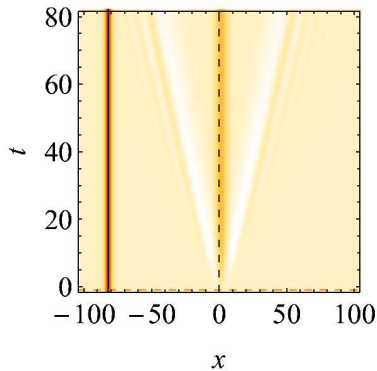}
    \caption{Numerical experiment portraying the trapped mode soliton formed at the center of the potential barrier for a stationary dark soliton positioned at $x_0 = -80$. The  external potential is indicated by the vertical dashed line. Parameters used are: $u_0=0.45$, $g_1=1/2,\,g_2=-1,\,V_0=0.1$, $\alpha=\sqrt{V_0}$.}
    \label{fig6}
\end{figure}

Furthermore, it is observed that radiation is emitted during the population of the trapped mode even for a stationary dark soliton and irrespective of its initial position from the potential. One of such trapped mode solitons appears at the center of the potential barrier, for an initial stationary dark soliton positioned at $x_0 = -80$, as displayed in Fig. \ref{fig6}. This trapped mode soliton is obtained for the initial dark soliton with an amplitude $u_0 = \sqrt{\frac{-2 V_0}{g_2}} = 0.45$ using the parameter setting, $V_0$ = 0.1, $g_1$ = 0.5 and $g_2= -1$. Although, the exact trapped mode soliton does not account for the radiation, however, in real situations such interactions are featured with background radiation that appears on either side of the potential barrier. The emergence of such radiations is due to the continuous interaction between the background of the dark soliton with the potential barrier.

\subsection{Effective position-dependent dispersion}
\begin{figure}[H]\centering
    \includegraphics[width=8cm,clip]{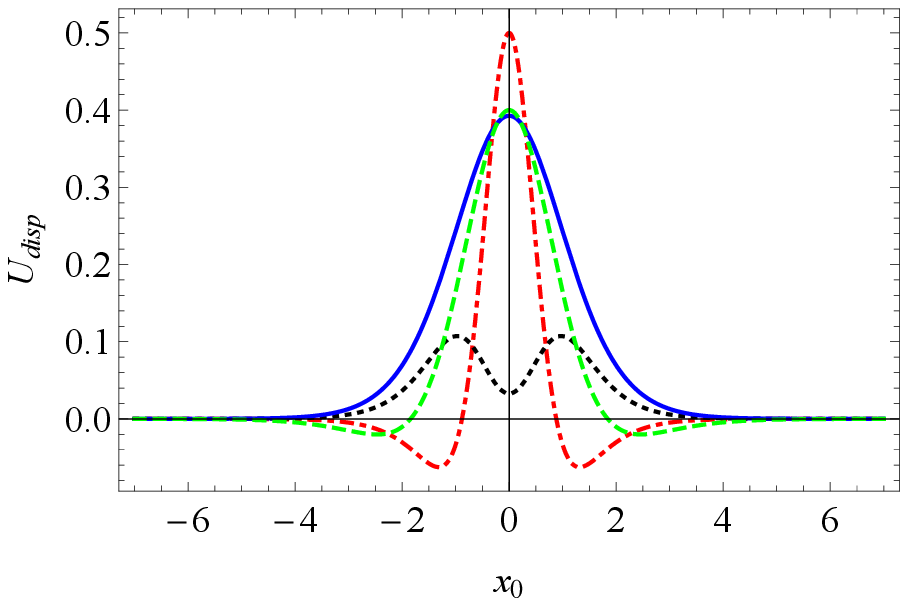}
    \begin{picture}(5,5)(5,5)
    \put(-24,140) {\color{black}{{\fcolorbox{white}{white}{\textbf{(a)}}}}}
    \end{picture}
    \includegraphics[width=8.11cm,clip]{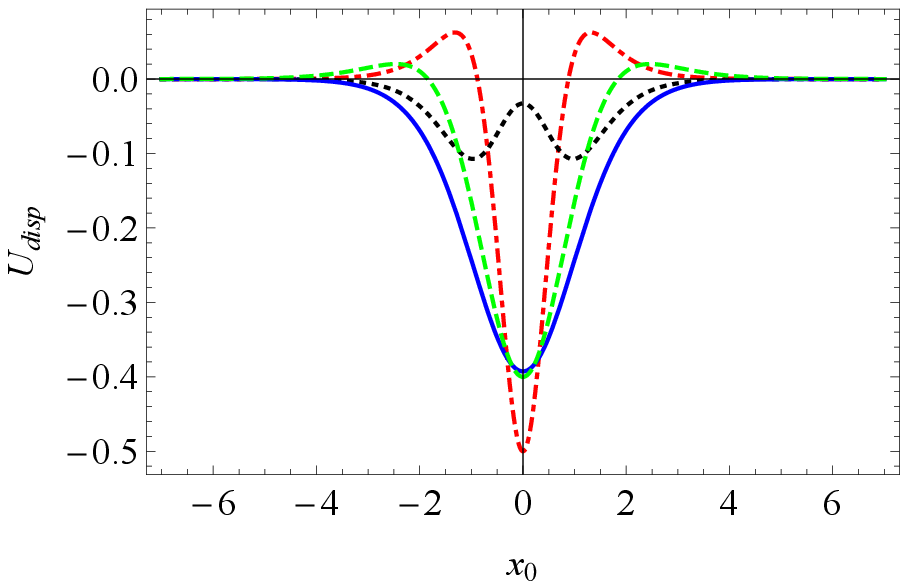}
    \begin{picture}(5,5)(5,5)
    \put(-27,50) {\color{black}{{\fcolorbox{white}{white}{\textbf{(b)}}}}}
    \end{picture}
    \caption{Effective dispersion potentials for different $x$-dependent dispersion functions $\delta(x)$, as given by Eq. (\ref{disps}). Dotted dashed red: Eq. (\ref{U1}), Dotted black: Eq. (\ref{U2}), Solid blue: Eq. (\ref{U3}), Dashed green: Eq. (\ref{U4}). (a) with bright soliton, $g_2=1$, and (b) with dark soliton, $g_2=-1$. Parameters used are: $U_0=1,\,g_1=1/2,\,n=2,\,w=1$.}
    \label{fig7}
\end{figure}

As an alternative method of achieving quantum reflection with dark solitons, we consider a different approach that can induce the same effect of an external potential. This involves the scattering of a dark soliton by an $x$-dependent dispersion. Such $x$-dependent dispersion can be assumed as any of the following forms of $\delta(x)$:
\begin{subequations}\label{disps}
    \begin{empheq}[left={\delta(x)=}\empheqlbrace]{align}
&
U_0\,\delta_{Dirac}(x/w),\label{disps1}\\\nonumber\\
& U_0\,\times\left\{\begin{array}{ll}1,\hspace{0.5cm}|x|\le w/2,\\0,\hspace{0.5cm}|x|>w/2,\end{array}\right.\label{disps2}\\\nonumber\\ &
U_0\,{\rm sech}(\alpha x),\label{disps3}\\\nonumber\\&U_0\,{\rm sech}^2(\alpha x).
\label{disps4}
    \end{empheq}
\end{subequations}
where $U_0$ and $w$ are the strength and width of the localised dispersion, respectively, and $\alpha=\sqrt{|U_0|}$. The effective potentials corresponding to these dispersion functions for the bright and dark solitons are calculated using Eq. (\ref{disps0}) as
\begin{subequations}
    \begin{empheq}[left={U_{disp}(x_0,0)=}\empheqlbrace]{align}
    &   \frac{g_1nU_0w\alpha^3\,{\rm sech}^4(\alpha x_0)}{4}[3-{\rm cosh}(2\alpha x_0)],\label{U1}\\\nonumber\\
    &   \frac{g_1n U_0 \alpha^2}{6}\left\{{\rm tanh}\left[\frac{\alpha}{2}(w-2x_0)\right]+{\rm tanh}\left[\frac{\alpha}{2}(w+2x_0)\right]\right\},\label{U2}\\\nonumber\\
    &   \pm\frac{g_1 n \pi U_0 \alpha^2}{8}{\rm sech}^4\left(\frac{\alpha x_0}{2}\right),\label{U3}\\\nonumber\\
    &   \frac{g_1 n  U_0 \alpha^2 x_0\,{\rm csch}^5\left(\alpha x_0\right)}{6}\left[27{\rm sinh}\left(\alpha x_0\right)+7{\rm sinh}\left(3\alpha x_0\right)-45\alpha{\rm cosh}\left(\alpha x_0\right)-3\alpha{\rm cosh}\left(3\alpha x_0\right)\right],\label{U4}
    \end{empheq}
\end{subequations}
where in the third line, the $+$ sign relates to the effective potential for the bright soliton and the $-$ sign corresponds to the dark soliton. For the other results, the effective potential is the same for both bright and dark solitons. In Fig. \ref{fig7}, we plot the effective dispersion potentials obtained above for both, the bright and dark solitons.
Since  the profile of the  effective dispersion  corresponding to  $\delta(x)=U_0\,{\rm sech}(\alpha x)$ has monotonically-decaying tails and without the appearance of pedestals besides the main peak, we select to focus on this function through our investigations below. Additionally, pulse profiles featured with pedestals result in complex scattering dynamics and eventually  lead to some further radiation during the numerical simulations.  A similar analogy to the case with external potential applies here; for dark solitons, quantum scattering requires the  effective potential to be a well. Thus, quantum scattering can be achieved with the parameter set $g_1>0$, $g_2<0$, $U_0>0$, while the parameter set $g_1>0$, $g_2<0$, $U_0<0$, leads to the classical scattering regime.\\

\begin{figure}[H]\centering
    \includegraphics[width=8cm,clip]{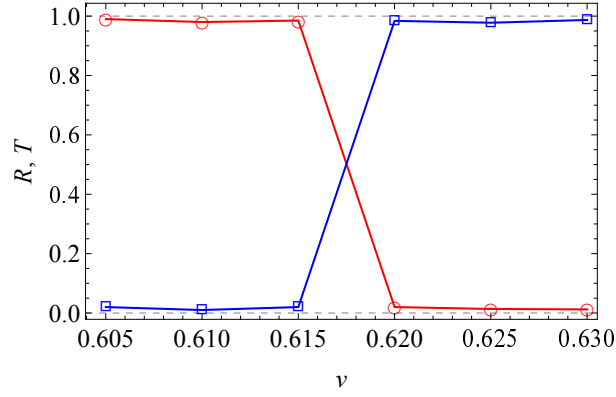}
    \caption{Reflection (red circles) and transmission (blue squares) coefficients versus initial speed of a dark soliton scattering by an $x$-dependent dispersion  (\ref{disps3}). Parameters used are: $U_0=0.4$, $g_1=1/2$, $g_2=-1$, $\alpha=\sqrt{g_1U_0}$, $x_0=-10$, $n=-4$.}
    \label{fig8}
\end{figure}

\begin{figure}[!h]\centering
    \includegraphics[width=5cm,clip]{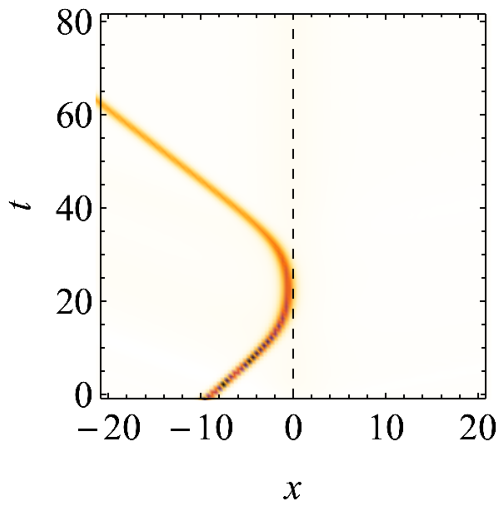}
    \begin{picture}(5,5)(5,5)
    \put(-30,133) {\color{black}{{\fcolorbox{white}{white}{\textbf{(a)}}}}}
    \end{picture}
    \includegraphics[width=5cm,clip]{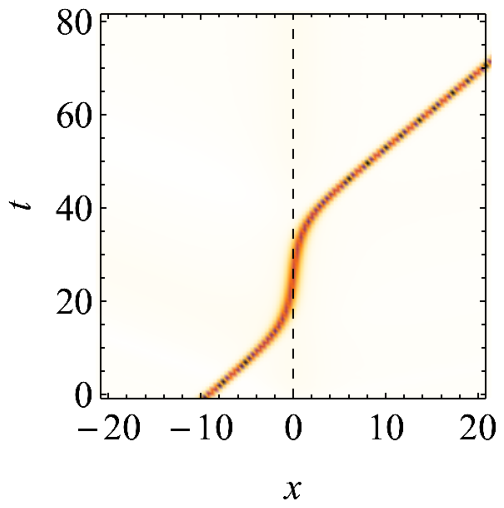}
    \begin{picture}(5,5)(5,5)
    \put(-30,133) {\color{black}{{\fcolorbox{white}{white}{\textbf{(b)}}}}}
    \end{picture}
    \includegraphics[width=5.15cm,clip]{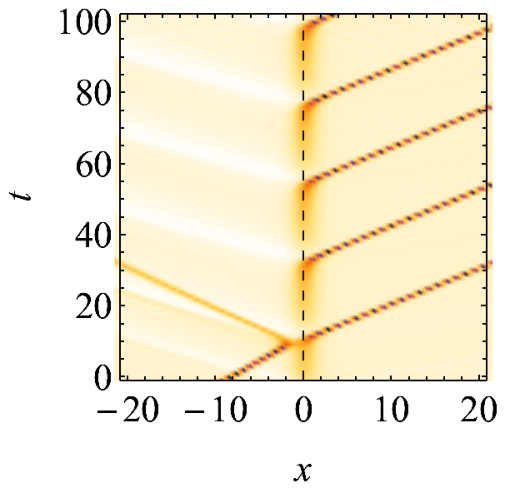}
    \begin{picture}(5,5)(5,5)
    \put(-100,126) {\color{black}{{\fcolorbox{white}{white}{\textbf{(c)}}}}}
    \end{picture}
    \caption{Quantum scattering of the dark soliton (\ref{ds})  by an $x$-dependent dispersion (\ref{disps3}). The  centre of the $x$-dependent dispersion is indicated be the vertical dashed line. (a) Quantum reflection with critical speed $v=v_c=0.615$, (b) transmission with $v=0.62$, and (c) multi-ejection $v=0.8$, $U_0=2$. Parameters used
        are the same of those in Fig. \ref{fig8}.}
    \label{fig9}
\end{figure}

\begin{figure}[H]\centering
    \includegraphics[width=8cm,clip]{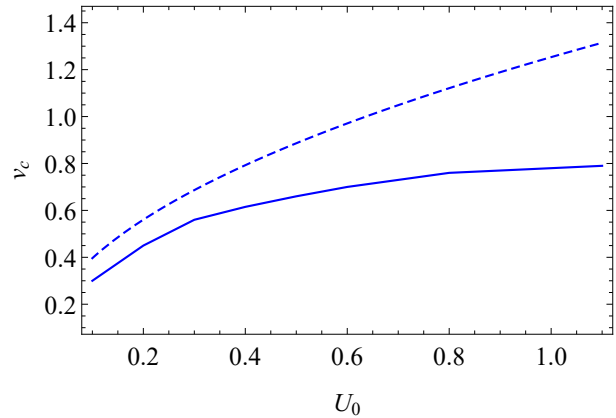}
    \caption{Critical soliton speed, $v_c$, at which the dark soliton will be
        reflected by the $x$-dependent dispersion (\ref{disps3}) in terms of its height, $U_0$. Solid curve is obtained from the numerical solution of Eq. (\ref{nlse2}), as explained in the text. Dashed curve is the result of time-dependent variational calculation (\ref{cspeed2}).  Parameters used are:  $g_1=1/2$, $g_2=-1$, $\alpha=\sqrt{g_1U_0}$, $x_0=-10$, $n=4$.}
    \label{fig10}
\end{figure}

In the following, we discuss the results obtained for quantum scattering parameters.
We solve numerically Eq. (\ref{nlse2}) using the function (\ref{disps3}). We set $V(x)=0$ and launch  $\psi_D(x,0)$ using Eq. (\ref{nlse2}) as an initial soliton. We defined the reflection and transmission  coefficients by Eqs. (\ref{r}) and (\ref{t}), where in this situation,  $l$ indicates the  length larger than the width of the $x$-dependent dispersion. In  Fig. \ref{fig8}, we plot the scattering coefficients versus the incident soliton speed.  The figure clearly shows the  possibility of obtaining quantum reflection, where a sharp transmission of the coefficients from full reflectance to full transmittance is achieved. This is confirmed by the spatio-temporal plots in Fig. \ref{fig9}. The critical speed for quantum reflection, $v_c$, is calculated numerically by running the initial soliton speed such that it gets trapped at the potential.  In Fig. \ref{fig10}, we plot the critial speed versus $U_0$. The dependence of $v_c$ on $U_0$ is found to be similar to that of dependence of $v_c$ on $V_0$ for the the external potential barrier case. Moreover, in Fig. \ref{fig9} (c), we show a multi-ejection behavior for the larger values of $U_0$, as the one observed for the external potential barrier.


\section{Scattering dynamics of dark solitons using time-dependent variational calculations }
\label{variationa}
In this section we use a time-dependent variational calculation to obtain an analytical account of the dynamical evolution of the scattering soliton.  The appearance of quantum reflection above a critical speed will be then confirmed. This will also provide analytical expressions of the critical speed for both cases of external potential and position-dependent dispersion.
\\\\
We employ the trial  function of the dark soliton in  (\ref{dstrial}) with varying the soliton centre  and velocity in $t$ and the use of the shifted intensity, namely, $v\rightarrow v(t)$, $x_0 \rightarrow x_0(t)$, and $\rho(x,t)=-(n\alpha/2)\,{\rm sech}^2[\alpha(x-x_0(t))]$.\\\\
The Lagrangian  function corresponding to the  NLSE, Eq. (\ref{nlse2}), then takes the form
\begin{eqnarray}\label{Lagr}
L&=&\int_{-\infty}^{\infty}i \psi^*(x,t)\psi_t(x,t) dx-\int_{-\infty}^{\infty}\left[g_1|\psi_x(x,t)|^2-\frac{1}{2}g_2|\psi(x,t)|^4+V(x)|\psi(x,t)|^2\right]dx\nonumber\\&=&\int_{-\infty}^{\infty}i \psi^*(x,t)\psi_t(x,t) dx-\left[E_0+U_{disp}(x_0,v)+U_{ext}(x_0,v)\right],
\end{eqnarray}
where we have  hidden the $t$-dependence for convenience.
\subsection{Effective external potential well}
In the presence of external potential barrier and with the absence of the $x$-dependent dispersion, $V(x)=V_0\,{\rm sech}^2(\alpha x)$, $\delta(x)=0$, the calculated Lagrangian in (\ref{Lagr}) with the trial  function of the dark soliton, (\ref{dstrial}), takes the form
\begin{eqnarray}\label{Lagr1}
L&=&-nv\dot{x_0}-\left\{-\frac{1}{3}g_1 n (\alpha^2+3v^2)-\frac{1}{6}g_2n^2\alpha-2nV_0{\rm csch}^2(\alpha x_0)\left[\alpha x_0{\rm coth}(\alpha x_0)-1\right]\right\}.
\end{eqnarray}
The corresponding Euler-Lagrange equations yield the following two equations of motion for the soliton centre and velocity
\begin{equation}\label{eom1}
 \dot{x_0}-2g_1v=0,
\end{equation}
\begin{equation}\label{eom2}
-n \dot{v}-6nV_0\alpha {\rm coth}(\alpha x_0){\rm csch}^2(\alpha x_0)+2V_0\alpha^2x_0(t){\rm csch}^4(\alpha x_0)\,[2+{\rm cosh}(2\alpha x_0)]=0.
\end{equation}
Integrating the above equation with respect to $x_0$ results in
\begin{equation}\label{inteom}
-2nV_0{\rm csch}^2(x_0)[\alpha{\rm coth}(\alpha x_0)x_0 -1]-\frac{1}{2}nv^2-c=0,
\end{equation}
where $c$ is the constant of integration.\\\\
For an initial position, $x_0(0)$, that is sufficiently far from the influence of the potential $V(x)$, the constant of integration can be calculated as
\begin{equation}\label{intcnst}
c=\frac{-nv_c^2}{2},
\end{equation}
where $v_c=v(0)$.\\\\
Substituting the obtained constant of integration in (\ref{inteom}) leads to
\begin{equation}\label{inteom1}
-2nV_0\,{\rm csch}^2(x_0)[\alpha\,x_0\,{\rm coth}(\alpha x_0) -1]+\frac{1}{2}n(v^2-v_c^2)=0.
\end{equation}
At the turning point which takes place at $t =t_c$, the velocity of the dark soliton vanishes, $v(t_c)=0$ and thus $v_c$ can be calculated from from the last equation as
\begin{equation}\label{cspeeda}
v_c=-2\sqrt{V_0}\left(\sqrt{\left\{x_0(0)\,\alpha\,{\rm coth}[x_0(0)\,\alpha]-1\right\}}\,{\rm csch}[x_0(0)\,\alpha]\right).
\end{equation}
The turning point position, $x_0(0)$, can be calculated from Eq. (\ref{eom2}) by setting the effective force to zero, namely $n\dot{v}=0$, which gives the solution $x_0(0)=0$. Direct substitution of $x_0(0)=0$ in the last equation leads to an undefined quantity. However, expanding it  in powers of  $x_0(0)$ and then taking the limit $x_0(0)\rightarrow0$, shows that all terms except the zeroth order one, vanish, and the result becomes
\begin{equation}\label{cspeed1}
v_c=2\sqrt{\frac{V_0}{3}}.
\end{equation}
This result is plotted by the dotted blue curve in Fig. (\ref{fig5}). While good agreement is obtained with the numerical values for smaller barrier heights, $V_0$, the deviation increases for larger barrier heights. The discrepancy is primarily due to not accounting for radiation and multi-ejection in the present variational approach.\\

\begin{figure}[H]\centering
    \includegraphics[scale=0.87,clip]{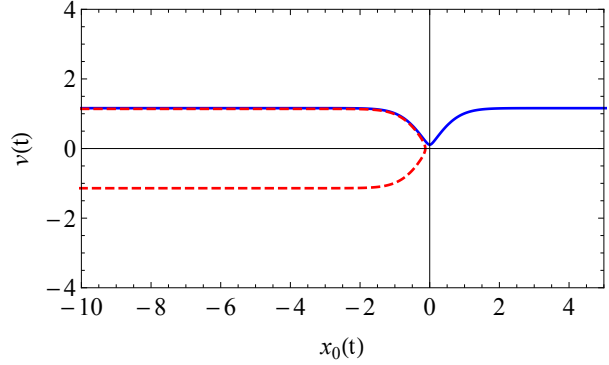}
    \caption{Dynamics of the variational parameters, speed $v(t)$ versus position $x_0(t)$ corresponding to NLSE with external potential, Eqs. (\ref{eom1}) and (\ref{eom2}). Sharp reflection (dashed red) occurs at the critical speed $v=v_c=1.15$ and transmission (blue) occurs at $v=1.156$. Parameters used are:  $V_0=1,\,g_1=1/2,\,g_2=-1,\,n=4$.}
    \label{fig11}
\end{figure}

In Fig. \ref{fig11}, we display the time varying position and speed obtained from the solutions of the equations of motion (\ref{eom1}) and (\ref{eom2}), that clearly show sharp transition from full reflection to full transmission.

\subsection{Effective well with position-dependent dispersion}
In the presence of position-dependent dispersion given by $\delta(x)=U_0\,{\rm sech}(\alpha x)$ and with $V(x)=0$,  the calculated Lagrangian in (\ref{Lagr}) with the trial  function of the dark soliton, (\ref{dstrial}), is reduced to
\begin{eqnarray}\label{Lagr2}
L&=&-nv\dot{x_0}-g_1\left[-\frac{1}{3}n\alpha^2-\frac{\pi}{4}nU_0v^2{\rm sech}^2\left(\frac{1}{2}\alpha x_0\right)-\frac{\pi}{8}nU_0\alpha^2{\rm sech}^4\left(\frac{1}{2}\alpha x_0\right)-nv^2-\frac{1}{6 g_1}g_2n^2\alpha\right],
\end{eqnarray}
where, based on our earlier discussion, we have neglected the nonhermitian term. The Euler-Lagrange equations yield the following two equations of motion for the soliton centre and velocity
\begin{equation}\label{eom3}
n \dot{x_0}-2g_1nv-\frac{\pi}{2}g_1nvU_0{\rm sech}^2\left(\frac{1}{2}\alpha x_0\right)=0,
\end{equation}
\begin{equation}\label{eom4}
-n \dot{v}+\frac{\pi}{4}g_1n\,\alpha\,U_0{\rm sech}^2\,\left(\frac{1}{2}\alpha x_0\right){\rm tanh}\left(\frac{1}{2}\alpha x_0\right)\left[v^2+\alpha^2{\rm sech}^2\left(\frac{1}{2}\alpha x_0\right)\right]=0.
\end{equation}
Following the derivation of $v_c$ expression (\ref{cspeed1}) in the previous section, we obtain the critical speed for the present case
\begin{equation}\label{cspeed2}
v_c=\frac{1}{2}\alpha\sqrt{\pi g_1U_0}.
\end{equation}
The comparative results for $v_c$ obtained through Lagrangian approach and those through numerical simulations are shown in Fig. \ref{fig10}. Similar to the previous case, deviation of the variational values from the numerical ones increase with increasing $U_0$, which, as we explained above is due not accounting for radiation. The trajectory in this case turns out to be very similar to that in Fig. \ref{fig11} confirming the critical scattering behaviour.

\section{conclusions and outlook}
\label{concsec} We have considered quantum reflection of dark
solitons in two setups, (i) in the presence of external potentials
and (ii) in the presence of an $x$-dependent dispersion. At the
outset, we have revisited the energy calculations and normalization
of both bright and dark solitons in order to determine the energy
functional and the condition that is necessary for quantum
reflection to occur with dark solitons. Both analytical and
variational calculations have shown that for quantum reflection of
dark solitons to occur in the aforementioned setups,  the actual
external potential or dispersion modulation need to be a
barrier-like function which leads to an effective potential well in
both cases.

In order to investigate the scattering dynamics, we have derived the
effective external potential corresponding to an external potential
and an $x$-dependent dispersion using a suitable trial function for
the dark soliton pulse. In the presence of an external potential, we
have considered the reflectionless P\"{o}schl-Teller potential and
for $x$-dependent dispersion we used the `sech' function, owing to
its corresponding effective potential without pedestals. This is
followed by the numerical and analytical study of the soliton
scattering.

Through our numerical investigations, we have observed quantum
reflection for dark solitons for both settings considered. The
critical speed required for such phenomenon tends to increase with
the height of the barrier. Moreover, our study revealed a transition
regime from clear quantum reflection to a multi-ejection behaviour
for larger barrier heights. Analytical calculation of the critical
speed in the case of external potential was also possible using the
exact trapped mode profile of the  dark soliton scattered by a
reflectionless potential barrier.

Then, we have derived the relation for the critical speed using
variational calculations for the two setups considered. The
variational study has successfully shown the sharp transition
behavior from full reflection to full transmission at the critical
speed. The deviation of the critical speed values calculated by the
exact trapped mode or by the variational calculation from the exact
numerical ones increases for larger potential strengths. This is due
to not including radiation in the analytical calculation. While the
inclusion of radiation complicates the calculations, we believe it
is worth investigating in a future work, as this will result in an
accurate analytical formula that predicts the threshold of quantum
reflection.

The above study is fully oriented to the quantum reflection, but in
general such interactions lead to many other interesting phenomena
such as  multi-ejection, snake trapping, and tree-ejection which are
possible by controlling the system parameters. In particular, we
have noticed a region in the parameters space where periodic
ejection of dark solitons is stimulated by the scattering. This
behavior will be considered in near future.

\section*{Acknowledgment}
The authors acknowledge the support of UAE University through Grants
No. UAEU-UPAR(1)-2019 and No. UAEU-UPAR(11)-2019.

\end{document}